\documentclass{paper}
\usepackage{geometry}
\usepackage[utf8]{inputenc}
\usepackage{graphicx}
\usepackage{tabularx}
\usepackage{cmbright} 
\usepackage{hyperref}

\title{Performance of the LHC injector chain after the upgrade and potential development}
\author{H.~Bartosik, G.~Rumolo }
\date{August 2020}

\begin{document}
\maketitle
\abstract{The CERN accelerator complex prepares protons for the four big experiments at the Interaction Regions of the Large Hadron Collider (LHC) as well as for a number of fixed target experiments, which take beam directly from the LHC injectors. Between 2010 and 2021, a series of major changes were designed and implemented to the injector complex under the LHC Injectors Upgrade (LIU) project, with the main goal to enable the injector chain to produce LHC beams with doubled intensity and brightness. As a byproduct of these upgrades, the fixed target beams are also expected to benefit and possibly access parameter ranges previously unexplored. This letter summarises the main fixed target beams produced by the LHC injectors, their expected performance after the upgrade for the various beam types and shows first results from 2021 beam commissioning.} 

\section{Introduction}

\begin{figure}[htb]
    \centering
    \includegraphics[trim=0 54 0 0, clip, width=0.64\textwidth]{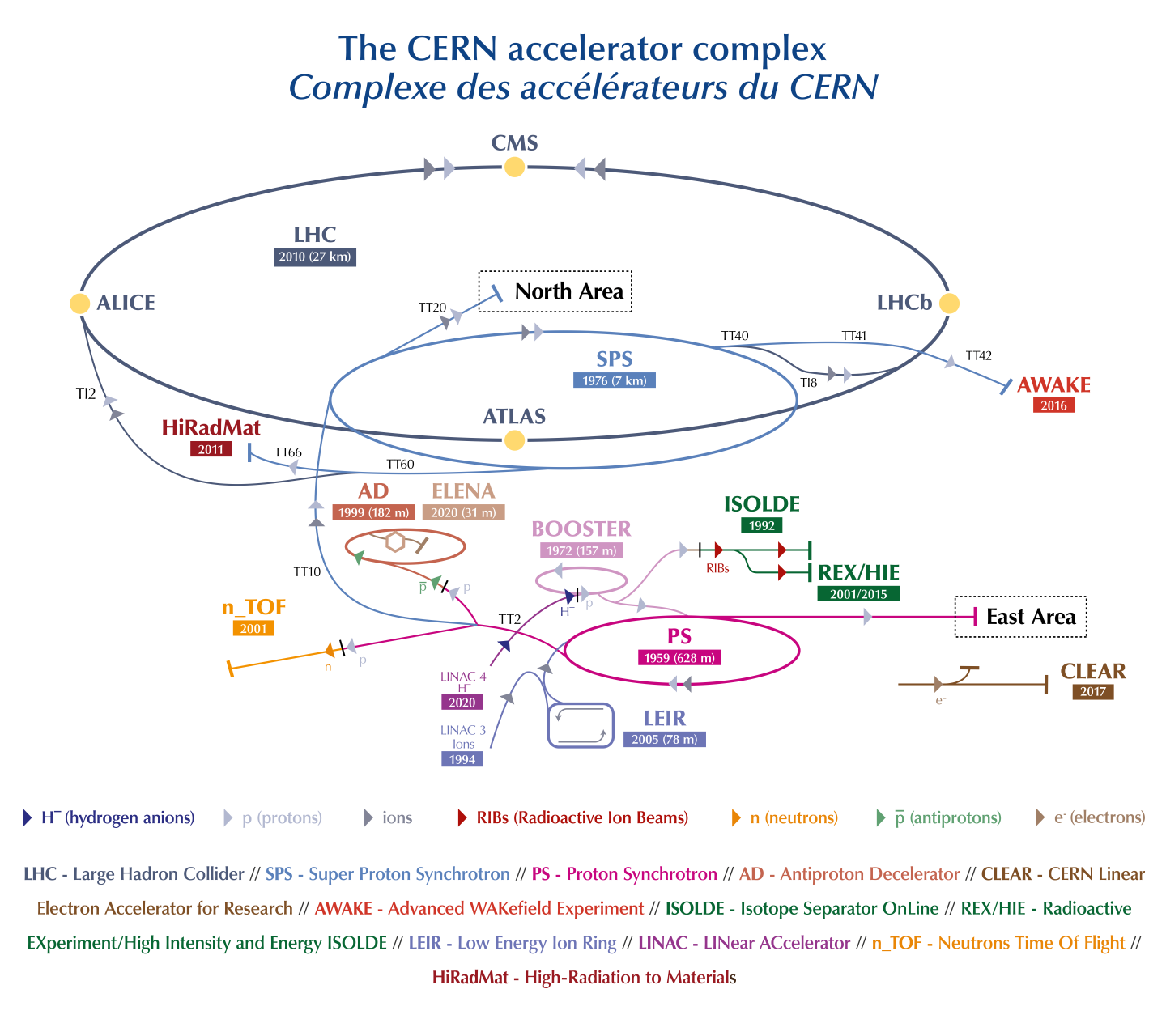}
    \caption{CERN accelerator complex and experimental facilities.}
    \label{fig:complex}
\end{figure}

The CERN accelerator complex does not only serve the experiments at the Large Hadron Collider (LHC), but it also offers a wide range of opportunities in the landscape of front-line physics. These include fixed-target (FT) experiments which aim to explore and understand novel phenomena in the field of particle physics, addressing some of today's outstanding questions with a complementary approach to the physics performed at the LHC. Different types of beams and experiments are required for this purpose and CERN offers the technologies and the infrastructure to carry out such experiments. CERN's physics program is therefore much broader and diverse, extending well beyond particle collider physics. The CERN injectors routinely provide beams at lower energies to facilities such as the Isotope Separator On Line DEvice (ISOLDE), the East Area (EA), the Antiproton Decelerator (AD) and Extra Low ENergy Antiproton (ELENA), the neutron Time-of-Flight facility (n-ToF), the High-Radiation to Materials (HiRadMat), the North Area (NA) and the Advanced WAKEfield Experiment (AWAKE).\\
The Physics Beyond Colliders (PBC) study~\cite{PBC1}, officially launched in 2016, aims at exploring the full scientific potential of the CERN accelerator complex and its scientific infrastructure through projects that are complementary to the LHC, High Luminosity LHC (HL-LHC) and other possible future circular colliders (FCC). The LHC injectors upgrade project (LIU) \cite{LIU} conceived, designed and implemented major hardware modifications in the whole injector complex, with the ultimate goal to improve the performance of the accelerator chain in the production of LHC beams. All non-LHC beams are expected to benefit from these upgrades, which also target improvements to the injector reliability and lifetime to cover the HL-LHC era until around 2040.\\
Figure~\ref{fig:complex} offers a schematics of the complex and the experimental facilities served by the different accelerators at CERN.

\subsection{Main upgrades of the accelerator complex under LIU}
Major upgrades of the whole injector complex were deployed during the LS2 to be able to fulfill the HL-LHC requirements. A new H- Linear Accelerator (Linac4) has replaced Linac2, which injected protons into the PSB until 2018. In the PSB a new injection region had to be installed to enable the H- charge exchange injection at 160 MeV instead of the 50 MeV multi-turn proton injection from Linac2. This has allowed doubling the beam brightness achievable in the PSB. Moreover, thanks to a new radiofrequency (RF) system and new main power supply, the beam energy at PSB extraction could also be increased from 1.4 GeV to 2 GeV for the beams sent to the PS. ISOLDE beams keep being extracted at 1.4 GeV, as the transfer line to ISOLDE and the ISOLDE dumps are not yet suited to cope with higher energy beams. For the extraction at 2 GeV, a new extraction septum was already installed during Run 2 and used at 1.4 GeV just before LS2.\\
The protons are now injected at 2 GeV into the PS, therefore allowing for brighter beams for the same space-charge tune shift. This required an upgrade of the injection region in the PS. Moreover, thanks to a dedicated longitudinal feedback system and the longitudinal impedance reduction of the 10 MHz RF cavities by about a factor of two, the threshold for the longitudinal coupled bunch instability has been significantly increased, such that the LIU baseline parameters can be comfortably achieved with the required beam quality.\\
In the SPS, reaching the LIU beam intensity required a major upgrade of the main 200 MHz RF system in combination with a combined impedance reduction/a-C coating campaign. A new beam dump system was also installed in long straight section 5 (LSS5) and new collimators will be installed in the transfer line to the LHC in order to cope with the doubled beam intensity and brightness. The full exploitation of the SPS upgrades is expected over the next couple of years, as the beam parameters of the LHC beams are gradually ramped up to the LIU target values.\\
Although all these upgrades were obviously aimed at improving the proton intensity and brightness of the LHC beams, an important impact on the future non-LHC beams can be also expected. Simulation and machine studies are now the key to assess the performance reach of the injector complex for the various facilities after LS2. Preliminary work was done before LS2 \cite{PBC2}, and some first results have been already seen in 2021 with the return to operation of the injectors complex after the LIU upgrade \cite{bartosik}.

\section{PSB}
The first synchrotron of the complex is the Proton Synchrotron Booster (PSB). As an outcome of the LIU project, the PSB is now successfully connected to the new H$^-$-linac (Linac4) and is able to accelerate proton beams from a kinetic energy of 160~MeV up to 2.0~GeV. The new energy range and the flexibility given by the H$^-$ injection has already proven to allow producing  beams about twice as bright as compared to before the upgrade. First studies on the intensity reach were conducted in 2021, however the focus was on recovering the pre-LS2 performance and demonstrating the capability of producing brighter beams as required for the LHC. Further studies on fixed target beams are expected as of 2022.
A short description of the main PSB beam types is provided below. 
\begin{itemize}
    \item The PSB beam can be extracted directly to the targets of the Isotope Separator On Line DEvice (ISOLDE), where radioactive nuclides are produced via spallation, fission, or fragmentation reactions. This is the highest intensity PSB beam, made up of 4 bunches extracted from each of the 4 PSB rings every 1.2~s, with intensities typically around $8\times 10^{12}$~p/bunch before LIU and a kinetic energy of 1.4~GeV. With the LIU upgrades, two potential developments have suddenly become available for this type of beam. On one hand, with a Linac4 current of 27~mA and the possibility to inject 150 turns per ring, the bunch intensity can be theoretically increased to about $1.6\times 10^{13}$~p/bunch, which would mean doubling the production rate on target. On the other hand, the kinetic energy at extraction can be increased to 2~GeV. Due to limitations in the transfer line to ISOLDE, in the dumps of the experimental area as well as in the proton current that can be sent on target to contain air activation (2~$\mu$A), the ISOLDE facility is not yet able to operationally use either of these options during the current run. Proposals for upgrade in LS3 and beyond are on the table, but there is no firm decision yet. In 2021, thanks to the connection to Linac4, it was possible  for the first time to accelerate to 1.4~GeV $1.2\times 10^{13}$~p in one ring (Ring 3) with losses along the cycle below 5\% (see Fig.\ref{fig:PSB-rec}, left plot). Machine studies to push this intensity beyond this value have not yet been conducted due to the uncertainty on the tolerable level of losses to avoid undesirable hot spots in the PSB, which may compromise interventions during technical stops or end-of-year shutdowns. Understanding the maximum intensity reach compatibly with an acceptable level of losses is the goal of 2022 machine studies, and beyond. These studies are crucial also in view of the possible upgrade of the Linac4 source and/or RFQ, which would potentially permit even higher intensities to be injected into the PSB.
    \item The PSB produces a single intense bunch to be injected into the PS for the n-TOF experiment with an intensity up to  $1\times 10^{13}$~p. This beam needs to be accelerated to 2~GeV. In the preparation of this beam in 2021, it was discovered that, above a certain intensity, the beam would suffer a strong horizontal instability at an energy between 1.4 and 2~GeV, against which the current feedback system was not enough to counteract. This horizontal instability could be successfully suppressed increasing linear coupling and it was demonstrated that an intensity of about $1.05\times 10^{13}$~p could be extracted at 2~GeV from the PSB with around 8\% losses (Fig.~\ref{fig:PSB-rec}, right plot). While this is sufficient to address the current requirements for the n-TOF experiment and further reduction of the losses can be reached by the improved resonance compensation schemes that were implemented later in the year (which can be even further optimised thanks to the additional power converters installed during the 2021-22 Year-End Technical Stop), it is important to understand the origin of this instability and determine at which intensity it practically limits the PSB operation for this type of beams.

 \begin{figure}[htb]
    \centering
    \includegraphics[trim=0 0 0 0, clip, width=0.74\textwidth]{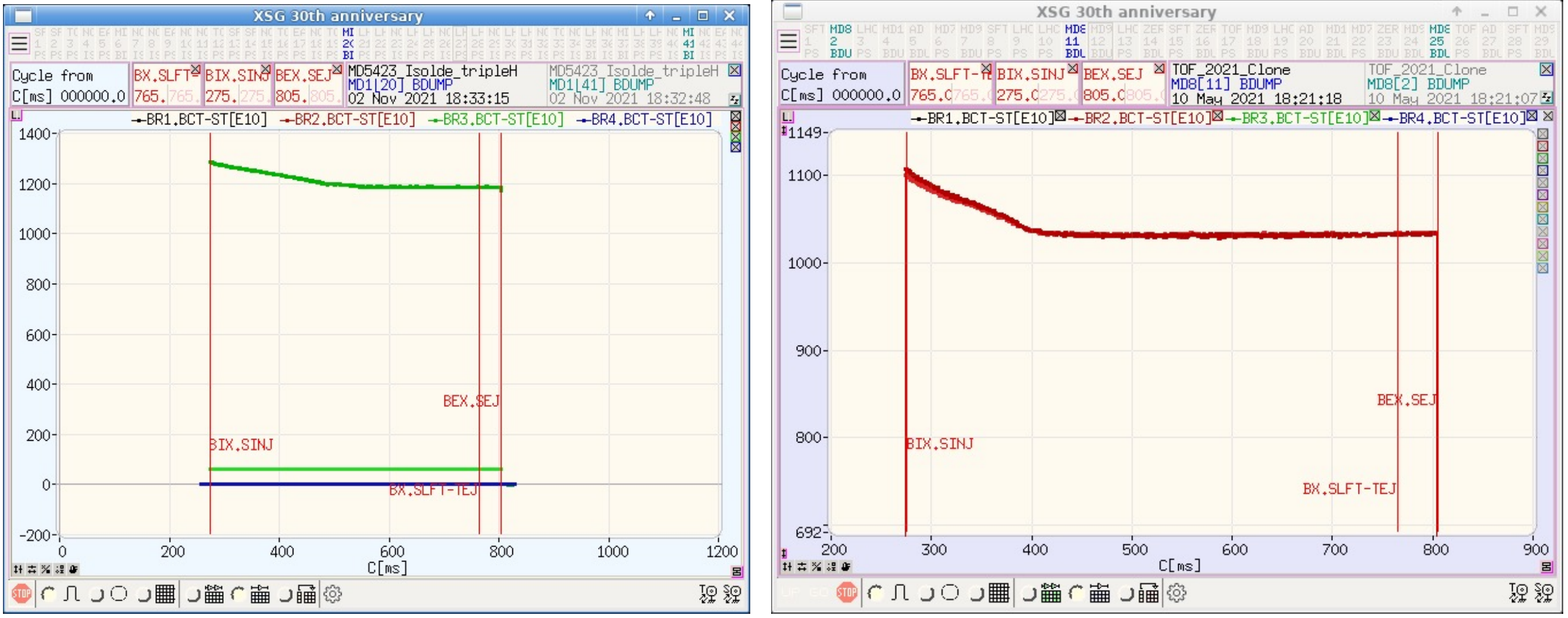}
    \caption{Record intensities accelerated in the PSB in 2021: $1.2\times 10^{13}$~p accelerated to 1.4~GeV in Ring 3 on an ISOLDE cycle (left) and $1.05\times 10^{13}$~p accelerated to 2~GeV in Ring 2 for n-TOF (right).}
    \label{fig:PSB-rec}
\end{figure}
    
        \item The beam used for fixed target physics in the SPS is usually referred to as MTE (as it undergoes Multi-Turn Extraction based on island trapping in the PS, \cite{MTE} and references therein) or SFTPRO beam in both the PSB and PS. This beam in the PSB reaches about $6\times 10^{12}$~p/ring (2 bunches per ring at extraction) and is required to feature large horizontal emittance (to facilitate the island splitting while remaining compatible with the aperture in the PSB, transfer lines and PS) and vertical emittance below 4~$\mu$m. This beam could be successfully produced in the PSB already in the first part of 2021 by using a combination of transverse painting and space charge blow-up at injection. The control and reproducibility of the beam parameters with the H$^-$ charge exchange injection from Linac4 is remarkable and the main visible result from the LIU upgrade for this type of beams.

\end{itemize}

\section{PS}
With the increased injection energy of 2.0~GeV and a modification of the longitudinal beam parameters at the PSB-PS transfer, the Proton Synchrotron (PS), next synchrotron in the chain, will be able to cope with the increased beam brightness. The PS accelerates beams to total energies ranging from 14 to 26~GeV. Thanks to the installation of a broadband Finemet cavity to act as a kicker for a longitudinal feedback system as well as the impedance reduction of the 10~MHz cavities, LHC beams are now longitudinally stable up to the intensities needed for LIU. 

\begin{figure}[htb]
    \centering
    \includegraphics[trim=0 0 0 0, clip, width=0.74\textwidth]{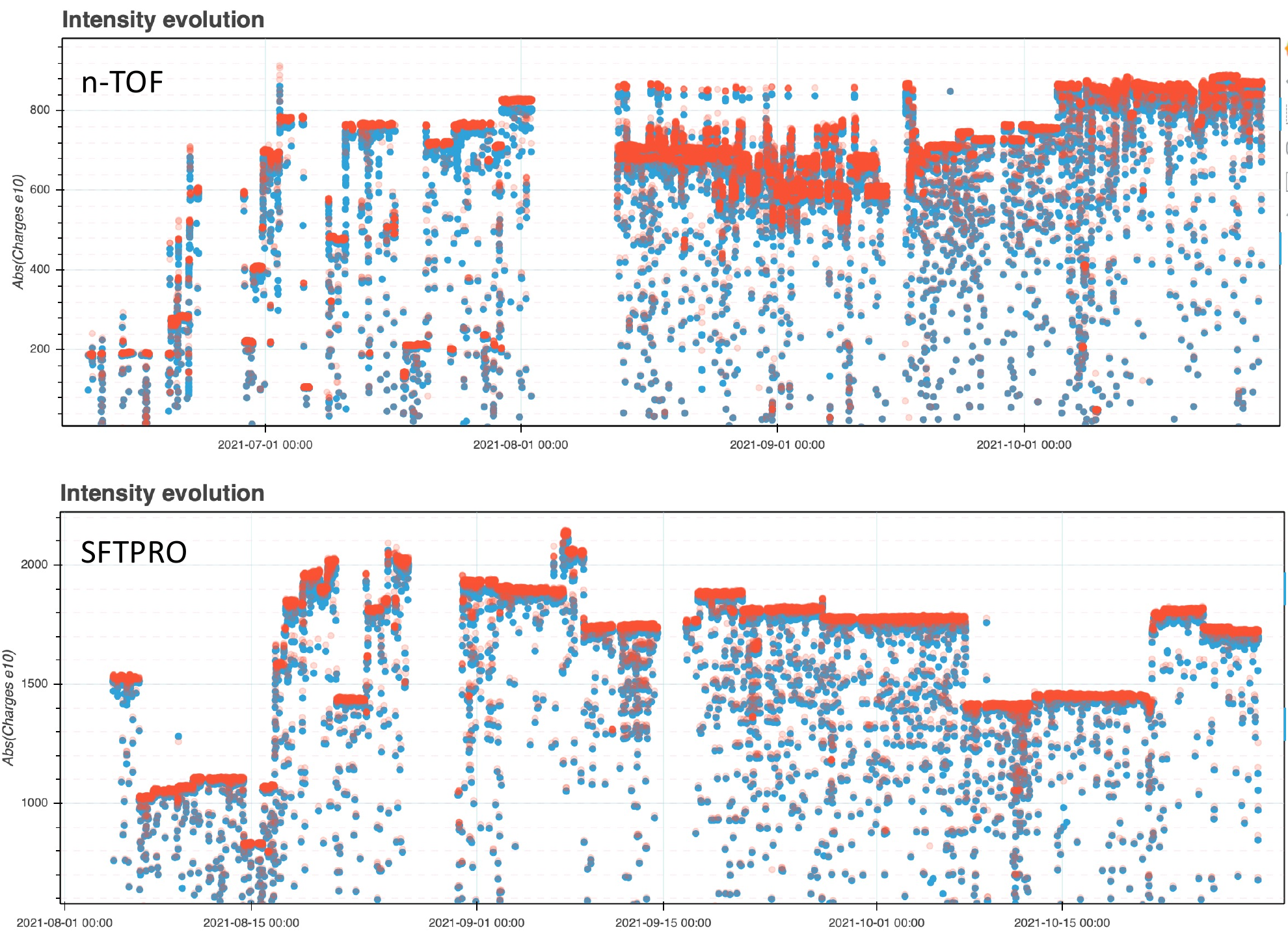}
    \caption{Extracted intensity to n-TOF target (top) and for SPS fixed target physics (SFTPRO, bottom) over the whole 2021 run.}
    \label{fig:PS-TOF-SFTPRO}
    \end{figure}

\begin{itemize}
    \item The PS can extract beam to the neutron Time-Of-Flight facility, n-TOF, which has been operating at CERN since 2001. nTOF is a pulsed neutron source coupled to a 200-metre flight path. It is designed to study neutron-nucleus interactions for neutron energies ranging from a few meV to several GeV. The n-TOF beam is made of one single short intense proton bunch, typically with $8\times 10^{12}$~p and fast rotated just before extraction in order to achieve 20~ns bunch length. In 2021, after the LIU upgrades, the beam was kept below $8\times 10^{12}$~p for most of the run, with values of $8.5\times 10^{12}$~p towards the end of the run (see Fig.~\ref{fig:PS-TOF-SFTPRO}, top graph). The plan is to take eventually $1\times 10^{13}$~p/bunch, compatibly with the PSB intensity reach, with the upgraded n-TOF target as well as with the known vertical instability at PS transition crossing.
    \item The proton beam coming from the PS can also be sent onto a target with a high yield of antiproton production in order to generate a secondary high-energy antiproton beam for the Antiproton Decelerator (AD) and Extra Low ENergy Antiproton (ELENA), a deceleration chain capable of turning it into a low-energy beam that can be used to produce antimatter. The AD beam in the PS is made of four bunches with about $3.5\times 10^{12}$~p/bunch spaced by about 100~ns and extracted at a total energy of 26~GeV. This beam has a large potential for intensity and/or brightness upgrade under LIU, which is however not being explored nor used due to other limitations coming from the target and the AD facility.
    \item The MTE (or SFTPRO) beam mostly operated around $1.8\times 10^{13}$~p in the PS over 2021 (see Fig.~\ref{fig:PS-TOF-SFTPRO}, bottom graph). One of the limitations of this beam is losses along the cycle and at extraction. Losses at extraction have been already reduced in the past by means of the implementation of the MTE with respect to the previous Continuous Transfer (CT) extraction, which would by construction lose a significant amount of protons on the septum blades. Residual losses at extraction still exist because of the continuous structure of the beam at the PS flat top and the extraction kicker rise time. These losses can be avoided if the beam is trapped in a barrier bucket having a gap synchronized with the extraction kicker rise time. In 2018, machine studies proved that the Finemet cavity installed under LIU can be  used for the capture of the SFTPRO beam in barrier bucket shortly before extraction, although no synchronization to SPS could be made due to the lack of time at flat top and hardware \cite{vadai}. In 2021 a new SFTPRO cycle was created with enough time at flat top to permit phase correction to synchronise the beam with the SPS. Besides, the necessary hardware was installed and connected to the LLRF chain. Machine studies conducted towards the end of the year showed that indeed the synchronisation with the SPS can be achieved. The final step next year will be the increase of intensity to $2.5-3\times 10^{13}$~p per PS cycle, which will require longitudinal stabilization and high intensity MTE on the longer flat top.

\end{itemize}

\section{SPS}
The largest machine of the injector complex is the Super Proton Synchrotron (SPS), which accelerates the beams to up to 450~GeV. The SPS main RF system (200~MHz) underwent a massive upgrade in power and controls under LIU. 
\begin{itemize}

    \item The SPS serves a number of experiments in the North Area (NA) by deploying a slow extraction with a spill length of 9.6~s (or higher) and then splitting the beam to different target stations. 
    This beam has typically a total intensity of about $3.5-4\times 10^{13}$~p (two injections from the PS) and a total energy of 400~GeV, which have remaned largely unchanged after LIU. 
    \item In the past (from 2007-2012), the SPS served the CERN Neutrinos to Gran Sasso (CNGS) experiment by fast extracting a high intensity proton beam. This beam featured an operational intensity of $4.0\times 10^{13}$~p per pulse and a record intensity of $5.3\times 10^{13}$~p per pulse in machine studies (CNGS+). Since the latter had the highest beam power per cycle ever produced at CERN so far, it is still kept as a reference for future development or installation of experiments requiring a CNGS-type beam. At the time of the studies, this beam was limited by losses both in  PS and SPS due to the large transverse emittances and to the limited SPS RF power. Thanks to LIU and the Multi-Turn Extraction (MTE) deployed at the PS-to-SPS  transfer, these limitations are expected to be relaxed in the future, making this intensity potentially available for routine operation. The proposed future experiment Search for Hidden Particles (SHiP) will use beam characteristics  similar to the operational CNGS, however deploying a slow extraction with 1~s spill length due to mechanical constraints coming from the target. As discussed above, higher beam intensities might be achievable for SHiP in the future similarly to CNGS. Due to the high duty cycle required for this experiment, losses during slow extraction have to be significantly reduced and are subject of detailed studies aiming at a loss reduction of more than a factor 4 compared to the presently achieved operational values \cite{SE-lossreduction}. More studies are planned on the SFTPRO beam in the SPS to experimentally quantify the benefit of LIU-RF upgrades for this type of beam, to operate it with the 800~MHz RF system active and use controlled longitudinal emittance blow-up to ensure longitudinal stability and to investigate the possibility to optimise transverse settings (optics, working point, transverse damper) to reduce losses at flat bottom or beginning of the ramp.
    \item Finally, the LHC beam in the SPS has been also used for fixed target studies at the High-Radiation to Materials (HiRadMat) facility. This is a users facility at CERN, designed to provide high-intensity pulsed beams to an irradiation area where material samples as well as accelerator component assemblies can be tested. The LHC beam after LIU will preserve the same time structure as before (4 trains of 72 bunches with 25~ns between bunches and 200~ns between trains), but it will have double intensity and brightness. This beam will be potentially available for HiRadMat if the facility will be able to swallow the stretched beam parameters.

\end{itemize}

\section{Summary and conclusions}
The LHC injectors serve a number of fixed target experiments by producing a wide range of beam parameters suited for the different applications. These beam parameters can indeed be improved after the implementation of the LIU upgrades in the injectors, and some benefits have been already visible in 2021 machine studies, like for example the achievement of record intensities on the ISOLDE and n-TOF beams in the PSB or the implementation of the barrier bucket to reduce the extraction losses of the MTE beam in the PS. Much still remains to be learnt, with machine studies planned to progress on this front in 2022 and beyond. Table~\ref{TAB:BeamParameters} provides an overview on the relevant beam parameters for the main fixed target beams produced in the LHC injectors complex, based on past experience and potential after LIU upgrades.
\newcolumntype{C}[1]{>{\centering\arraybackslash}p{#1}}
\begin{table}[h]
\begin{center}
\caption{Overview of beam types available at the CERN injector complex after LIU.}
\begin{tabular}{lC{1.3cm}C{1.3cm}C{1.3cm}C{1.3cm}C{1.3cm}C{1.3cm}|C{1.3cm}}
\hline
 & ISOLDE & n-TOF & AD & SHiP & NA & LHC & CNGS+   \\
\hline
Total Energy [GeV] & 2.4/3.0 & 20 & 26 & 400 & 400 & 450 & 400 \\
Total intensity [$1\times10^{13}$ p] & 6.4 & 1.0 & 1.40 & 4.0 & 4.0 & 6.7 & 5.3 \\
Cycle length [s] & 1.2 & 1.2 & 2.4 & 7.2 & 10.8 & 21.6 & 6.0\\
Beam power [kW] & 20/26 & 27 & 24 & 356 & 237 & 223 & 566 \\
Total bunch length [ns] & 230/200 & 20 & 38 & - & - & 1.6 & 4 \\
Number of bunches  & 4 & 1 & 4 & coasting & coasting & 288 & 4200 \\
Bunch spacing [ns] & 572 & - & 100 & - & - & 25 & 5 \\
Extraction type & fast & fast & fast & slow & slow & fast & fast \\
\hline
\label{TAB:BeamParameters}
\end{tabular}
\end{center}
\end{table}


\end{document}